\documentclass[aps,pre,reprint,superscriptaddress,amsmath,showpacs,float]{revtex4-1}
\usepackage[hyperindex,breaklinks,colorlinks=true]{}
\usepackage{graphicx,sidecap}

\begin{document}
\title{Universality classes of the generalized epidemic process on random networks}

\author{Kihong Chung}
\affiliation {Department of Physics,
Korea Advanced Institute of Science and Technology, Daejeon
34141, Korea}

\author{Yongjoo Baek}
\affiliation {Department of Physics, Technion, 
Haifa 32000, Israel}

\author{Meesoon Ha}
\email[]{msha@chosun.ac.kr}
\affiliation{Department of Physics Education, Chosun University,
Gwangju 61452, Korea}

\author{Hawoong Jeong}
\email[]{hjeong@kaist.edu}
\affiliation{Department of Physics and Institute for the
BioCentury, Korea Advanced Institute of Science and Technology,
Daejeon 34141, Korea}

\date{\today}

\begin{abstract}
We present a self-contained discussion of the universality classes of the generalized epidemic process (GEP) on Poisson random networks, which is a simple model of social contagions with cooperative effects. These effects lead to rich phase transitional behaviors that include continuous and discontinuous transitions with tricriticality in between. With the help of a comprehensive finite-size scaling theory, we numerically confirm static and dynamic scaling behaviors of the GEP near continuous phase transitions and at tricriticality, which verifies the field-theoretical results of previous studies. We also propose a proper criterion for the discontinuous transition line, which is shown to coincide with the bond percolation threshold.
\end{abstract}



\maketitle

\section{Introduction} \label{sec:intro}

Percolation provides a useful paradigm for understanding universal properties of contagions. Studies of dynamical percolation showed that simple contagions with immunizing effects, whose transmission events are independent of each other, exhibit a continuous phase transition belonging to the bond percolation universality class on both regular lattices~\cite{Cardy1985,*Janssen1985} and random networks~\cite{Newman2002}. Recently there has been a surge of interest in percolation models with cooperative effects, which may exhibit discontinuous transitions and associated tricritical behaviors~\cite{Bizhani2012,Araujo2014}. In the context of contagion, cooperative effects are present when the probability of infection changes according to the number of contacts with infected individuals, as is often the case with {\it social contagions} of behaviors~\cite{Centola2010}. Thus we expect that simple percolation models with cooperative effects would give us much insight into the universal properties of social contagions.

The generalized epidemic process (GEP)~\footnote{Janssen {\it et al.}~\cite{Janssen2004} originally called the model generalized general epidemic process (GGEP), but we choose a shorter name GEP following the suggestion of \cite{Bizhani2012}.}, independently proposed by \cite{Janssen2004} and \cite{Dodds2004,*Dodds2005}, is a simple variant of dynamical percolation with cooperative effects. The model features an infection probability which changes according to the number of contacts with infected individuals. It has been shown that both continuous and discontinuous transitions are possible on regular lattices~\cite{Janssen2004,Bizhani2012} and random networks~\cite{Dodds2004,*Dodds2005,Bizhani2012,KChung2014}, with scaling properties belonging to the bond percolation universality class~\cite{Janssen2004,Bizhani2012,KChung2014} in the vicinity of critical points. Intermediate tricritical behaviors were also studied by a field-theoretical approach, whose notable differences from the ordinary bond percolation behaviors include the change of the upper critical dimension from $6$ to $5$ and the breakdown of symmetry between the percolation probability and the giant cluster size~\cite{Janssen2004}.

However, even in the mean-field (MF) case, our knowledge of the GEP remains incomplete. First, discussions of finite-size effects at tricriticality have been missing. Since actual contagions and numerical simulations involve finite systems, the development of a finite-size scaling (FSS) theory is crucial for a complete and verifiable description of tricritical behaviors. Second, in the case of random networks with a single infected seed, the discontinuous transition line has not been properly identified. It was claimed that a discontinuous transition occurs when the self-consistency equation for the order parameter abruptly develops a nonzero double root~\cite{Bizhani2012}. 
However, this criterion does not guarantee the actual occurrence of a transition, because an infinite infected cluster corresponding to the nonzero root may not be achievable with a finite probability starting from a single infected seed.
To address this problem, we need to find a criterion which ensures that a finite infected cluster is able to expand indefinitely with a nonzero probability~\cite{*[{We note that for nonequilibrium systems discontinuous transitions can be defined in different ways depending on the initial state. Ours correspond to {\it spinodal transitions} defined in }] [] Janssen2016}.

In this work, we present a self-contained treatment of the GEP on Poisson random networks. The rest of this paper is organized as follows: Sec.~\ref{sec:model} gives a description of the model; Sec.~\ref{sec:critical}, based on the locally tree-like structure of Poisson random networks, presents exact derivations of the epidemic threshold for all types of transitions and the MF exponents for continuous transitions; in Sec.~\ref{sec:fss}, using these analytical results, we propose a set of FSS hypotheses for (tri)critical behaviors of various physical quantities, all of which are numerically verified; finally, we summarize our findings and conclude in Sec.~\ref{sec:summary}.

\section{Model} \label{sec:model}

We consider a Poisson random network~\cite{Erdos1959,*Gilbert1959} of $N$ nodes randomly connected by links so that the mean degree (number of neighbors) of each node is $c$. A node is in one of the following four states: unexposed ($\mathbf{S_1}$), exposed ($\mathbf{S_2}$), infected ($\mathbf{I}$), and removed ($\mathbf{R}$). At each instant, a randomly chosen $\mathbf{I}$ node tries to infect every susceptible ($\mathbf{S_1}$ or $\mathbf{S_2}$) neighbor and is promptly removed ($\mathbf{I} \to \mathbf{R}$) from the dynamics. An $\mathbf{S_1}$ neighbor becomes infected ($\mathbf{S_1} \to \mathbf{I}$) with a probability $\lambda$, or becomes exposed ($\mathbf{S_1} \to \mathbf{S_2}$). An $\mathbf{S_2}$ neighbor is infected ($\mathbf{S_2} \to \mathbf{I}$) with a probability $\mu$, or nothing happens. The case $\lambda = \mu$ corresponds to the ordinary dynamical percolation, also called the susceptible--infected--removed (SIR) model. On the other hand, in social contagions exposure usually increases the chance of infection ($\lambda < \mu$)~\cite{Centola2010}. Nevertheless, for the sake of generality, we do not put any constraint on the ranges of $\lambda$ and $\mu$. After each update {~\footnote{{The asynchronous updating scheme is employed in our simulation. However, our theory is expected to be exact for other updating schemes ({\it e.g.,} synchronous ones) as well, provided that the network is locally tree-like, only a single infection attempt is made through each link, and a random tiebreaker is assumed for any simultaneous infection attempts on a single target node.}}}, the time is increased by $1/I(t)$, where $I(t)$ is the number of $\mathbf{I}$ nodes at time $t$.

The contagion starts by putting all nodes in the $\mathbf{S_1}$ state, and infecting a single seed node at random. It ends when all $\mathbf{I}$ nodes vanish, leaving a cluster of $\mathbf{R}$ nodes. In the limit $N \to \infty$, if this cluster is infinitely large and accounts for a finite fraction $r$ of the network, an {\it epidemic outbreak} is said to occur. It turns out that the probability $P_\infty$ of an epidemic outbreak becomes nonzero only if $\lambda$ is above a certain threshold $\lambda_c$ determined by the other parameters. This phenomenon, called {\it epidemic transition}, can also be interpreted as a percolation transition with the percolation probability given by $P_\infty$ and the giant cluster size given by $r$. The outbreak size $r$ can be regarded as an order parameter.

\section{Phase diagram \\ and critical behaviors} \label{sec:critical}

\subsection{Outbreak size}

In order to describe the dependence of $r$ on $\lambda$ and $\mu$, it is useful to approximate the network as a tree, which is composed of a series of levels~\cite{Gleeson2008,*Gleeson2009}. Each level consists of an infinite number of nodes, and links exist only between adjacent levels, so that nodes in the $n$-th level pass the contagion on to their neighbors in the ($n+1$)-th level. The fraction of $\mathbf{R}$ nodes in the $n$-th level is denoted by $r_n$. For the SIR model, $r_n$ satisfies a recursive relation~\cite{Newman2002,Bizhani2012}
\begin{align} \label{eq:self_consistency_sir}
r_{n+1} = 1 - \sum_{k=0}^{\infty} p_k [1 - r_n + (1-\lambda)r_n]^k,
\end{align}
where $p_k$ is the degree distribution given by
\begin{equation}\label{eq:p_k}
p_k = \frac{c^k}{k!}e^{-c},
\end{equation}
{where the mean degree is $\langle k \rangle = c$.}
For the GEP, we need a substitution~\cite{Bizhani2012}
\begin{equation}
(1-\lambda)^m \to (1-\lambda)(1-\mu)^{m-1}
\end{equation}
for every term with $m \ge 2$, which changes Eq.~\eqref{eq:self_consistency_sir} to
\begin{align} \label{eq:self_consistency_gep}
r_{n+1} &= 1- \sum_{k = 0}^{\infty}p_k \left[\frac{1-\lambda}{1-\mu}(1-\mu r_n)^k - \frac{\mu-\lambda}{1-\mu}(1-r_n)^k \right]\nonumber\\
&= 1 - \frac{1-\lambda}{1-\mu}e^{-cr_n\mu} + \frac{\mu-\lambda}{1-\mu}e^{-cr_n} \equiv f(r_n).
\end{align}
This relation defines a map $f$. If the initial seed belongs to the zeroth level, we have $r_0 = 0$, so $r = r_\infty$ must be the stable fixed point accessible from $r_0 = 0$. The location of this fixed point is determined by the behaviors of $f$ near $r = 0$, which can be expanded as
\begin{align} \label{eq:r_self_consistency}
f(r) = &\lambda cr + [(1-\lambda)\mu-\lambda] \frac{(cr)^2}{2} \nonumber\\
&\quad + \frac{(1-\lambda)\mu^3 - \mu + \lambda}{1-\mu}\frac{(cr)^3}{6} + O(r^4).
\end{align}

\begin{figure}[]
\includegraphics[width=0.95\columnwidth]{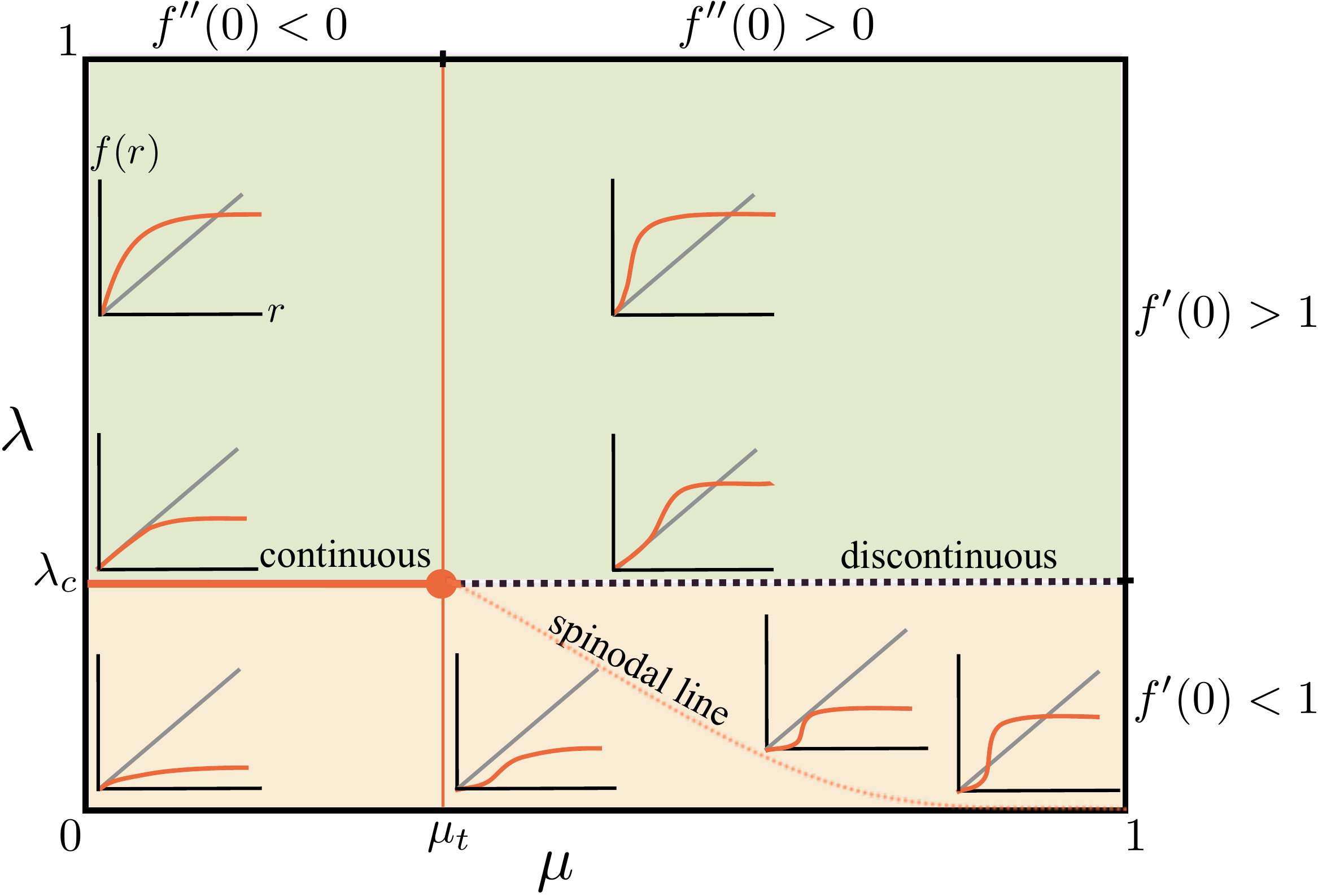}
\caption{\label{fig:fig1} (Color online) The schematic phase diagram of the GEP on Poisson random networks is drawn by two control parameters, which represent the first infection probability $\lambda$ and the second infection probability $\mu$, respectively. Insets are schematic illustrations of the fixed-point condition $f(r)=r$ in the corresponding region.}
\end{figure}

The insets of Fig.~\ref{fig:fig1} are schematic illustrations of Eq.~\eqref{eq:self_consistency_gep}, where curves representing the map $r_{n+1} = f(r_n)$ are shown together with the diagonal line $r_{n+1} = r_n$ representing the fixed-point condition. The coefficient of $r$ in Eq.~\eqref{eq:r_self_consistency}, which is obtained from $f'(0)$, shows that $r = 0$ becomes an unstable fixed point for $\lambda > 1/c$, making the adjacent fixed point with $r > 0$ stable. This implies that the epidemic threshold is always located at $\lambda = \lambda_c \equiv 1/c$, which agrees with the bond percolation threshold~\cite{Barrat2008,Callaway2000,Newman2001}.

The transition nature can be determined from the coefficient of $r^2$ in Eq.~\eqref{eq:r_self_consistency}, which is obtained from $f''(0)$. The coefficient is negative for $\mu \le \mu_t \equiv 1/(c-1)$, which shows that the transition of $r$ is continuous in this range of $\mu$. For $0 < \varepsilon \equiv (\lambda - \lambda_c)/\lambda_c \ll 1$, $r$ scales like
\begin{equation}
r \sim \varepsilon^\beta,\quad \beta = \begin{cases}
1 & \text{ for $0 \le \mu < \mu_t$},\\
1/2 & \text{ for $\mu = \mu_t$}.
\end{cases}
\end{equation}

Meanwhile, if $\mu > \mu_t$, the transition of $r$ becomes discontinuous. In this case, Eq.~\eqref{eq:r_self_consistency} has a stable fixed point at $r > 0$ even for $\lambda < \lambda_c$; but here $r = 0$ is by itself a stable fixed point, and there is an unstable fixed point between the two stable fixed points, so this nonzero stable fixed point becomes inaccessible. Thus, when the contagion starts from a single seed, one should not mistake the lower boundary of this region for the discontinuous transition line. By analogy with equilibrium discontinuous transitions, the boundary can be more aptly called the ``spinodal line,'' as marked in Fig.~\ref{fig:fig1}.

We have thus obtained a full phase diagram of the GEP on Poisson random networks, which is illustrated in Fig.~\ref{fig:fig1} in terms of $r$. Now we discuss the (tri)critical behaviors of the system in the vicinity of continuous transitions for $\mu \le \mu_t$.
\begin{table}
\centering \caption{\label{tab} Mean-field (tri)critical exponents of the GEP.}
\begin{tabular}{ccccccc}
\hline\hline
& $~~~~\beta~~~~$ & $~~~~\beta'~~~~$ & $~~~~\tau~~~~$ & $~~~~\bar{\nu}~~~~$ & $~~~~\nu_{_{\parallel}}~~~~$\\
\hline
$~~~~\mu < \mu_t~~~~$ & $1$ & $1$ & $5/2$ & $3$ & $1$\\
$\mu = \mu_t$ & $1/2$ & $1$ & $5/2$ & $5/2$ & $1$\\
\hline\hline
\end{tabular}
\end{table}

\subsection{Outbreak probability}

If $P_R$ denotes the probability of an $\mathbf{R}$ cluster of size $R$, $P_\infty$ can be obtained from
\begin{equation} \label{eq:p_infty}
P_\infty = 1 - \sum_{R = 1}^\infty P_R.
\end{equation}

A Poisson random network is {\it locally tree-like}: it hardly contains any loops of finite length~\cite{Bollobas1985}. For $\mu$ to take effect, there must be two different chains of infections leading to a single node: one for exposure, and the other for infection. This is virtually impossible for finite $\mathbf{R}$ clusters due to the rarity of finite loops. Thus, $\sum_{R = 1}^\infty P_R$ on the right-hand side of Eq.~\eqref{eq:p_infty} is affected only by $\lambda$, and as a result the behaviors of $P_\infty$ are indistinguishable from the MF bond percolation behaviors shown by the SIR model on Poisson random networks~\cite{Newman2002}. We note that a similar mechanism was also found in a related model of cooperative co-epidemics~\cite{WCai2015}. Hence, for $\varepsilon \ll 1$, $P_\infty$ shows a critical behavior
\begin{equation} \label{eq:betap}
P_\infty \sim \varepsilon^{\beta'}, \quad \beta' = 1.
\end{equation}
Thus the identity $\beta = \beta'$, which reflects the symmetry between the percolation probability and the giant component size of bond percolation, breaks down at tricriticality. This is consistent with the field-theoretical prediction of \cite{Janssen2004}. Besides, $P_R$ exhibits an algebraic decay
\begin{equation}
P_R \sim R^{1-\tau}, \quad \tau = 5/2
\end{equation}
at $\varepsilon = 0$, which is another property of the MF bond percolation. We note that $\beta'$, and $\tau$ can be derived exactly by generating function techniques~\cite{Barrat2008,Callaway2000,Newman2001,Newman2002,Cohen2002}. An interested reader is referred to Appendices~\ref{app:beta} and \ref{app:tau}.

\subsection{Exponent $\bar{\nu}$}

We present a heuristic argument for the value of the exponent $\bar{\nu}$, which describes the divergence of the correlation volume $\xi_{\rm v}\sim |\varepsilon|^{-\bar{\nu}}$. From percolation theory~\cite{Stauffer1991}, for $|\varepsilon| \ll 1$ one expects $P_R \sim R^{1-\tau} e^{-R/R_0}$, where the cutoff size $R_0$ satisfies a scaling relation $R_0 \sim |\varepsilon|^{-1/\sigma}$. This implies
\begin{equation}
P_\infty \simeq 1 - \int_{1}^{R_0} \mathrm{d}R P_R \sim R_0^{2-\tau} \sim \varepsilon^{(\tau-2)/\sigma},
\end{equation}
which yields a known scaling relation~\cite{Stauffer1991}
\begin{equation}
\beta' = \frac{\tau - 2}{\sigma}.
\end{equation}
Finite-size effects are important when the outbreak size is comparable to the cutoff size, {\it i.e.}, $Nr \simeq R_0$. To put it another way, finite-size effects are felt when $\xi_{\rm v}$ is comparable to $N$. Thus $\bar{\nu}$ is obtained as
\begin{equation} \label{eq:nu_bar}
\bar{\nu} = \beta + \frac{1}{\sigma} = \beta + \frac{\beta'}{\tau - 2} = \begin{cases}
3 & \text{ for $\mu < \mu_t$},\\
5/2 & \text{ for $\mu = \mu_t$}.
\end{cases}
\end{equation}
Since Poisson random networks are infinite-dimensional, we have $\bar{\nu} = d_u \nu$~\cite{Hong2007}, where $d_u$ is the upper critical dimension and $\nu$ has the MF value, $\nu_{_{\rm MF}}=1/2$~\cite{Stauffer1991}. Therefore, Eq.~\eqref{eq:nu_bar} implies that $d_u$ changes from $6$ to $5$ at the tricritical point, which agrees with the field-theoretical result~\cite{Janssen2004}.

\subsection{Dynamical exponent $\nu_{_\parallel}$}

The above discussions show that the second infection probability $\mu$ is irrelevant to the dynamics as long as the $\mathbf{R}$ cluster remains finite. Since the contagion starts from a single seed, no infinite $\mathbf{R}$ clusters can be formed in a finite time $t$. Therefore, the dynamical exponent must be given by $\nu_{_\parallel} = 1$, which is equal to the same exponent of the MF bond percolation~\cite{Cardy1985,Janssen1985}.

For Poisson random networks, $\nu_{_\parallel} = 1$ can be justified by a simple heuristic argument. We note that any node reached by following a randomly chosen link has on average $c$ additional neighbors~\cite{Bollobas1985}. Thus, at any finite time $t$, each $\mathbf{I}$ node tries to spread the disease to $c$ susceptible neighbors on average.The average number of $\mathbf{I}$ nodes therefore evolves like
\begin{equation}
\langle I(t+1) \rangle = \langle I(t) \rangle c\lambda = \langle I(t) \rangle \lambda/\lambda_c = \langle I(t) \rangle (1+\varepsilon).
\end{equation}
For $\varepsilon \ll 1$, this relation has an approximate solution $\langle I(t) \rangle \sim e^{t/t_0} \sim e^{\varepsilon t}$. Hence the characteristic time scale of the system satisfies
\begin{equation}
t_0 \sim \varepsilon^{-\nu_{_\parallel}}, \quad \nu_{_\parallel} = 1.
\end{equation}

All (tri)critical exponents obtained so far are summarized in Table~\ref{tab}.

\section{FSS forms and numerical results} \label{sec:fss}

\subsection{Extended FSS for the surviving probability}

In what follows, we propose an extended FSS theory for finite size $N$ and finite time $t$~\cite{ZBLi1995,*CChoi2013,*MJLee2014}, which is useful for numerical verifications of all exponents obtained so far. We postulate that the surviving probability $P_\mathrm{s}$, defined as the probability that the network has at least one $\mathbf{I}$ node, scales as
\begin{equation} \label{eq:fss_surv}
P_\mathrm{s}(t, N, \varepsilon) = b^{y_\mathrm{s}} \, P_\mathrm{s}(b^{\bar{z}}t,  bN, b^{-1/\bar{\nu}}\varepsilon),
\end{equation}
where $b$ is a scaling factor and $\bar{z} \equiv \nu_{_{\parallel}}/\bar{\nu}$. In the limit $t \to \infty$ and $N \to \infty$, we have $P_\mathrm{s} \to P_\infty$. This is consistent with Eq.~\eqref{eq:betap} only for $y_\mathrm{s} = \beta'/\bar{\nu}$. Thus, choosing $b = 1/N$, we can reduce Eq.~\eqref{eq:fss_surv} to
\begin{align} \label{eq:reduced_fss_surv}
P_\mathrm{s}(t,N,\varepsilon) &= N^{-\beta'/\bar{\nu}} \Psi_\mathrm{s}(t N^{-\bar{z}}, \varepsilon N^{1/\bar{\nu}}) \nonumber\\
&= t^{-\beta'/\nu_{_\parallel}} \Phi_\mathrm{s}(t N^{-\bar{z}}, \varepsilon N^{1/\bar{\nu}}),
\end{align}
where $\Phi_\mathrm{s}(x_1,x_2) \equiv \Psi_\mathrm{s}(x_1,x_2)/x_1^{\beta'/\nu_{_\parallel}}$. We note that the validity of the FSS forms like Eqs.~\eqref{eq:fss_surv} and \eqref{eq:reduced_fss_surv} is logically independent of our analytical results obtained in Sec.~\ref{sec:critical}. By numerical means, we aim to verify both FSS forms and the values of critical exponents.

\begin{figure}[b]
\includegraphics[width=0.9\columnwidth]{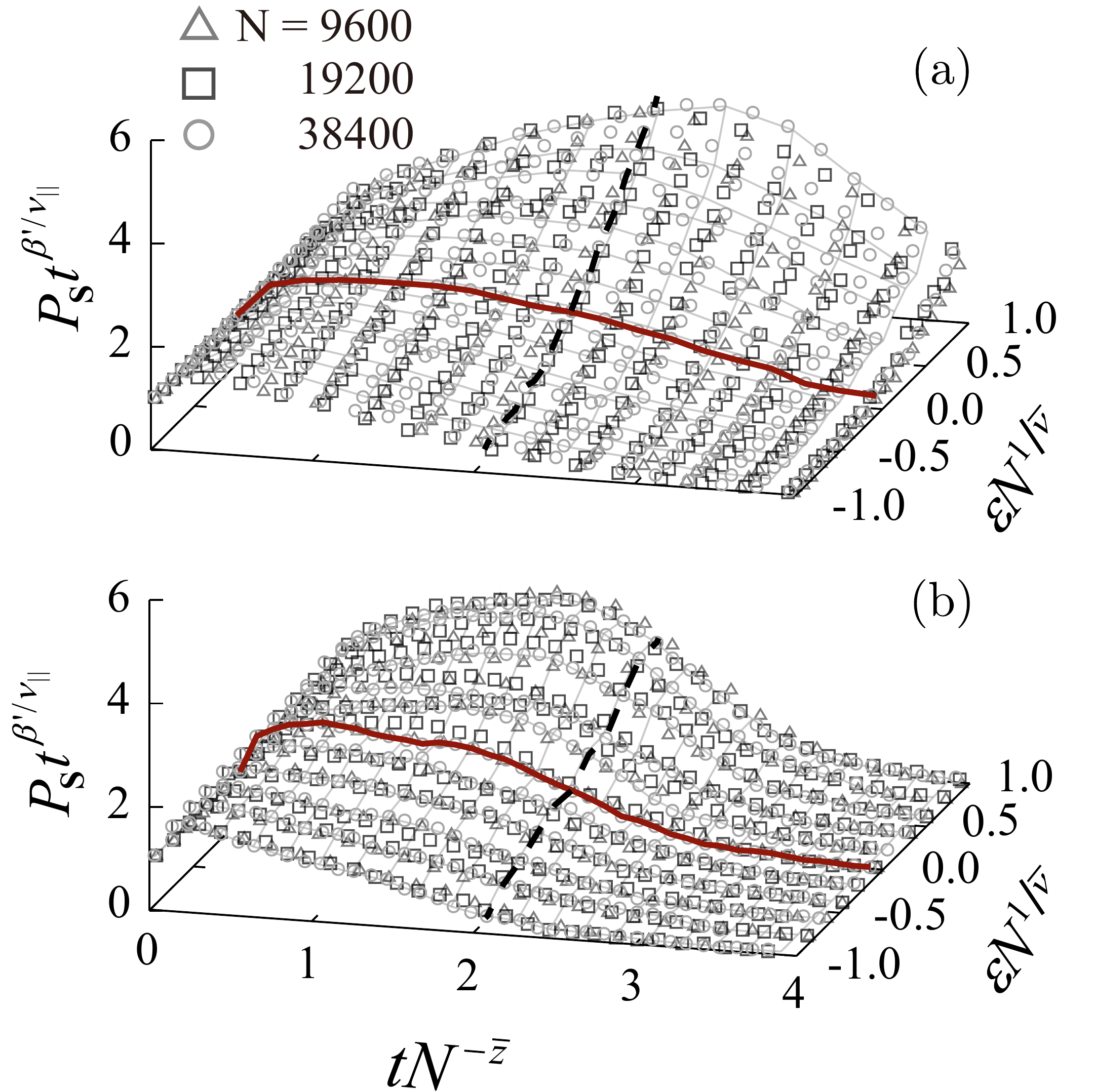}
\caption{\label{fig:fig2} (Color online) Numerical verifications of the extended FSS form $\Phi_\mathrm{s}$ for $P_\mathrm{s}$ on Poisson random networks with $\langle k \rangle = 5$: (a) $\mu = 0.21 < \mu_t$ (critical) and (b) $\mu = 0.25 = \mu_t$ (tricritical), where cross sections $\varepsilon N^{1/\bar{\nu}} = 0$ and $tN^{-\bar{z}} \approx 2$ are highlighted by solid (red or gray) and dashed (black) lines. The exponent values of Table~\ref{tab} are used with $\bar{z} = \nu_{_{\parallel}}/\bar{\nu}$.}
\end{figure}

\begin{figure*}[]
\includegraphics[width=0.95\textwidth]{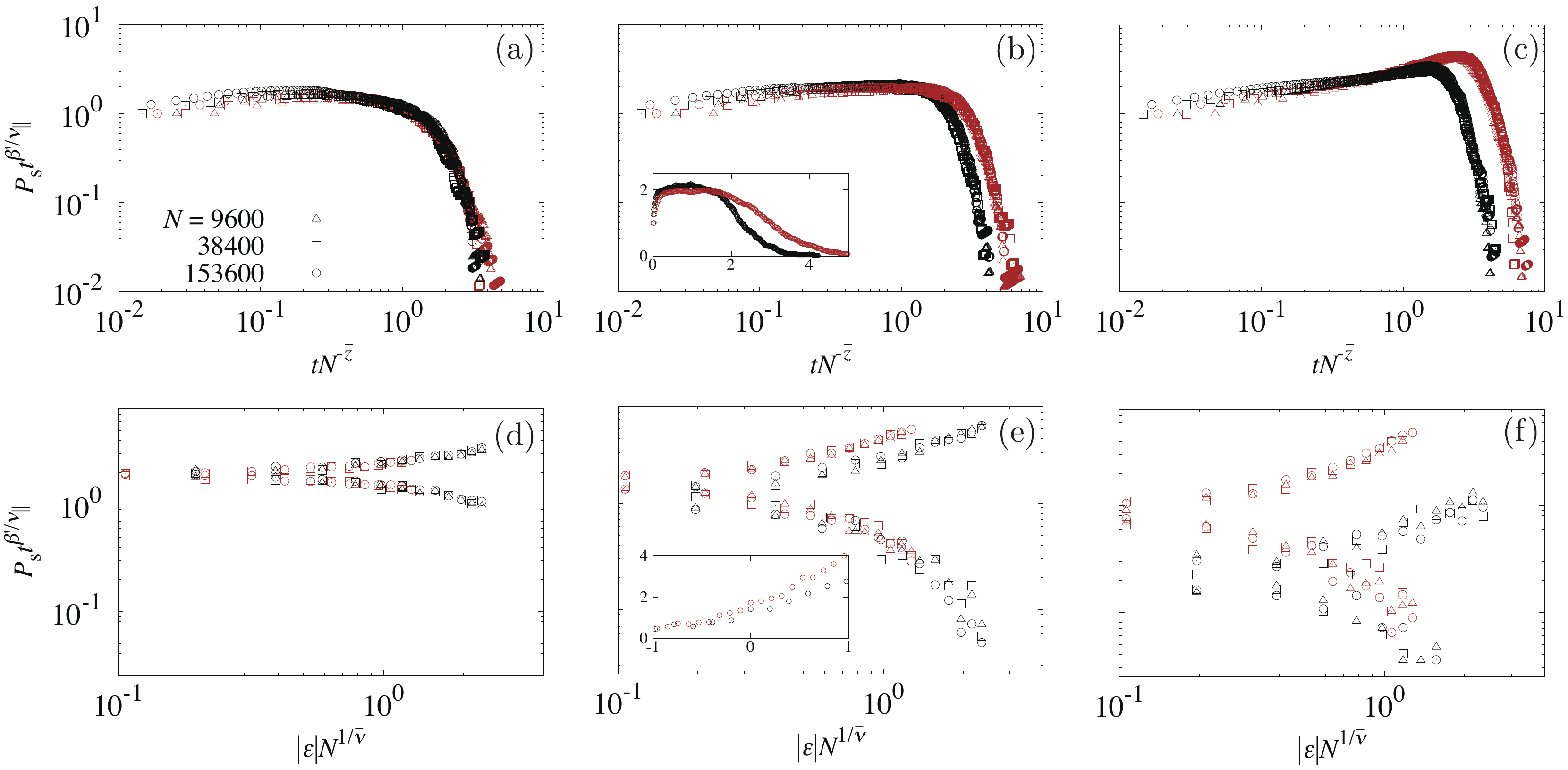}
\caption{\label{fig:fig3} (Color online) Cross sections of data collapse for $P_\mathrm{s}$ in Fig.~\ref{fig:fig2} at $\mu=0.21$ (red or gray) and at $\mu=0.25$ (black) are plotted in the double-logarithmic scales. In the upper panel, the values of $\epsilon N^{1/\bar{\nu}}$ are fixed at (a) $-1.0$, (b) $0.0$, and (c) $1.0$ as $t N^{-\bar{z}}$ varies, while in the lower panel, the values of $t N^{-\bar{z}}$ are fixed at (d) $0.5$, (e) $2.0$, and (f) $3.0$ as $\epsilon N^{1/\bar{\nu}}$ varies. The insets of (b) and (e) represent the data of $N=153600$ in real scales, respectively.}
\end{figure*}

The FSS form of Eq.~\eqref{eq:reduced_fss_surv} can be confirmed by a three-dimensional data collapse, as shown in Fig.~\ref{fig:fig2}. Also see Fig.~\ref{fig:fig3} for a closer view of several cross sections including those marked by solid (red or gray) and dashed (black) lines. All results with respective exponents are consistent with the scaling forms.

\subsection{Extended FSS for the density of $\mathbf{I}$ nodes}

\begin{figure}[]
\includegraphics[width=0.925\columnwidth]{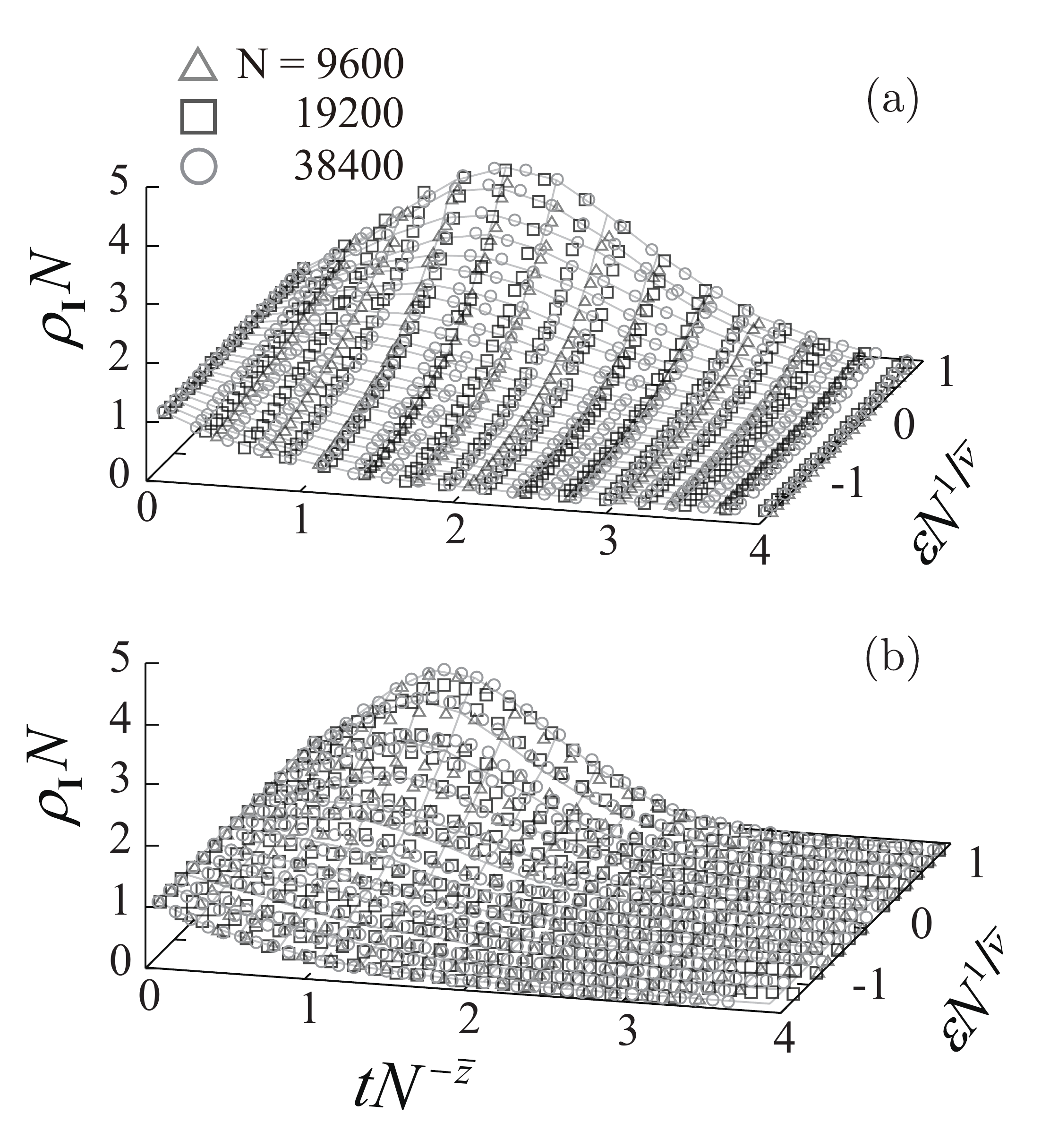}
\caption{\label{fig:fig4} (Color online) Numerical verifications of the extended FSS form $\Phi_\mathbf{I}$ for $\rho_\mathbf{I}$ on Poisson random networks with $\langle k \rangle = 5$: (a) $\mu = 0.21$ (critical) and (b) $\mu = 0.25$ (tricritical).}
\end{figure}

An extended FSS theory can also be formulated for the density of $\mathbf{I}$, which is denoted by $\rho_{_\mathbf{I}}$. We postulate that
\begin{align} \label{eq:rhoR_EFSS}
\rho_{_\mathbf{R}}(t,N) = b^{y_\rho} \rho_{_\mathbf{R}}(b^{\bar{z}}t,bN,b^{-1/\bar\nu}\varepsilon),
\end{align}
wehre $\rho_{_\mathbf{R}} \equiv \langle R \rangle/N$  is the mean density of $\mathbf{R}$ nodes.
In the limit $N \to \infty$, we expect $\rho_{_\mathbf{R}} \simeq r P_\infty \sim \varepsilon^{\beta+ \beta'}$ for $\varepsilon \ll 1$, which implies $y_\rho = (\beta+\beta')/\bar\nu$. Moreover, the average density of $\mathbf{I}$ nodes satisfies $\rho_{_\mathbf{I}} = \partial_t \rho_{_\mathbf{R}}$. Thus, we can write a scaling form for $\rho_{_\mathbf{I}}$ as
\begin{align}
\rho_{_\mathbf{I}}(t,N,\varepsilon) = b^{(\beta+\beta')/\bar\nu + \bar{z}} \rho_{_\mathbf{I}}(b^{\bar{z}}t,bN,b^{-1/\bar\nu}\varepsilon).
\end{align}
From Table~\ref{tab}, it is easy to see that
\begin{equation}
(\beta+\beta')/\bar\nu + \bar{z} = 1.
\end{equation}
Thus, putting $b = 1/N$, we obtain a scaling form
\begin{align} \label{eq:I_time}
\rho_{_\mathbf{I}}(t,N,\varepsilon) = N^{-1} \Phi_\mathbf{I}(t N^{-\bar{z}},\varepsilon N^{1/\bar{\nu}}).
\end{align}
As shown in Fig.~\ref{fig:fig4}, this scaling form is applicable to both critical and tricritical points.

\begin{figure}[]
\includegraphics[width=0.9\columnwidth]{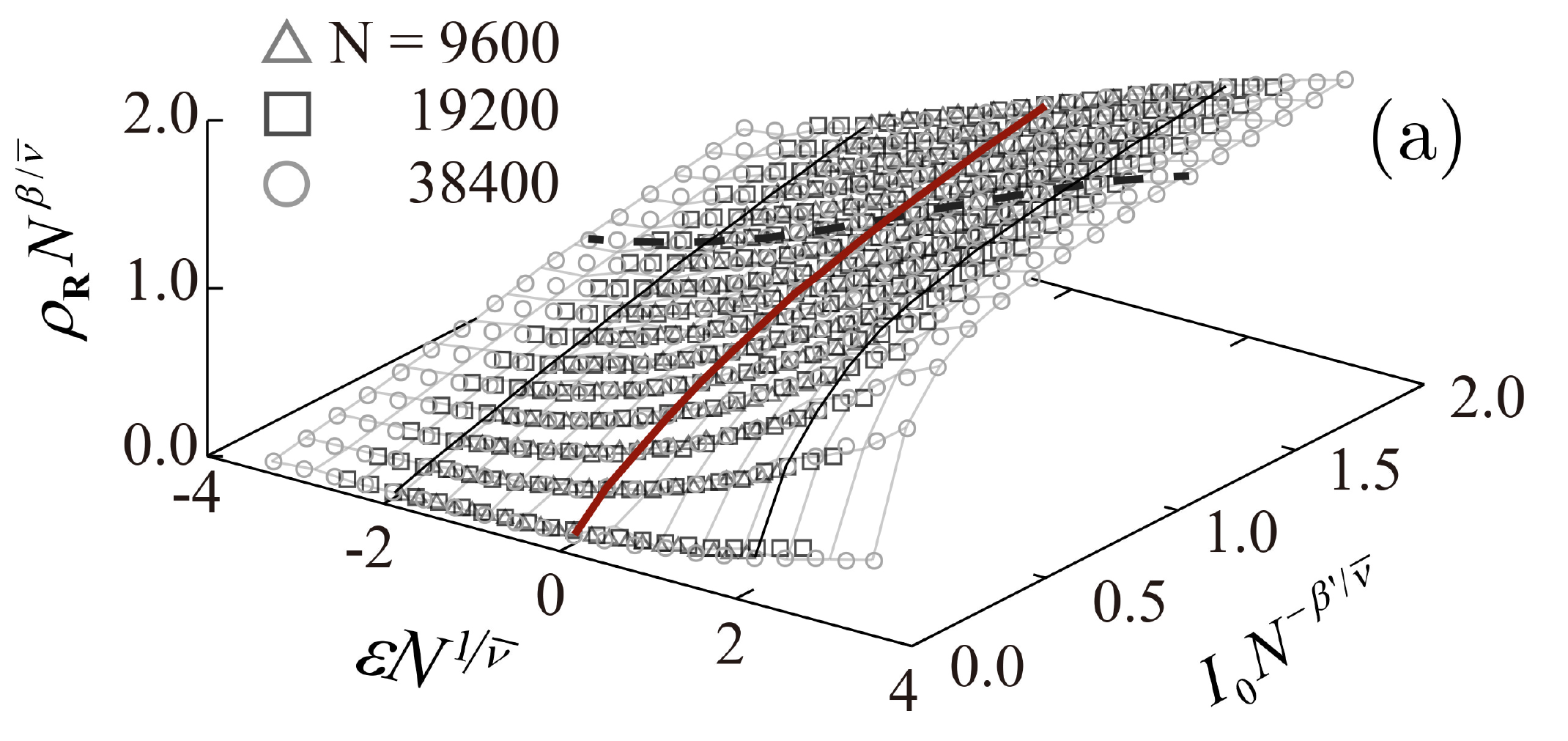}\\
\includegraphics[width=0.85\columnwidth]{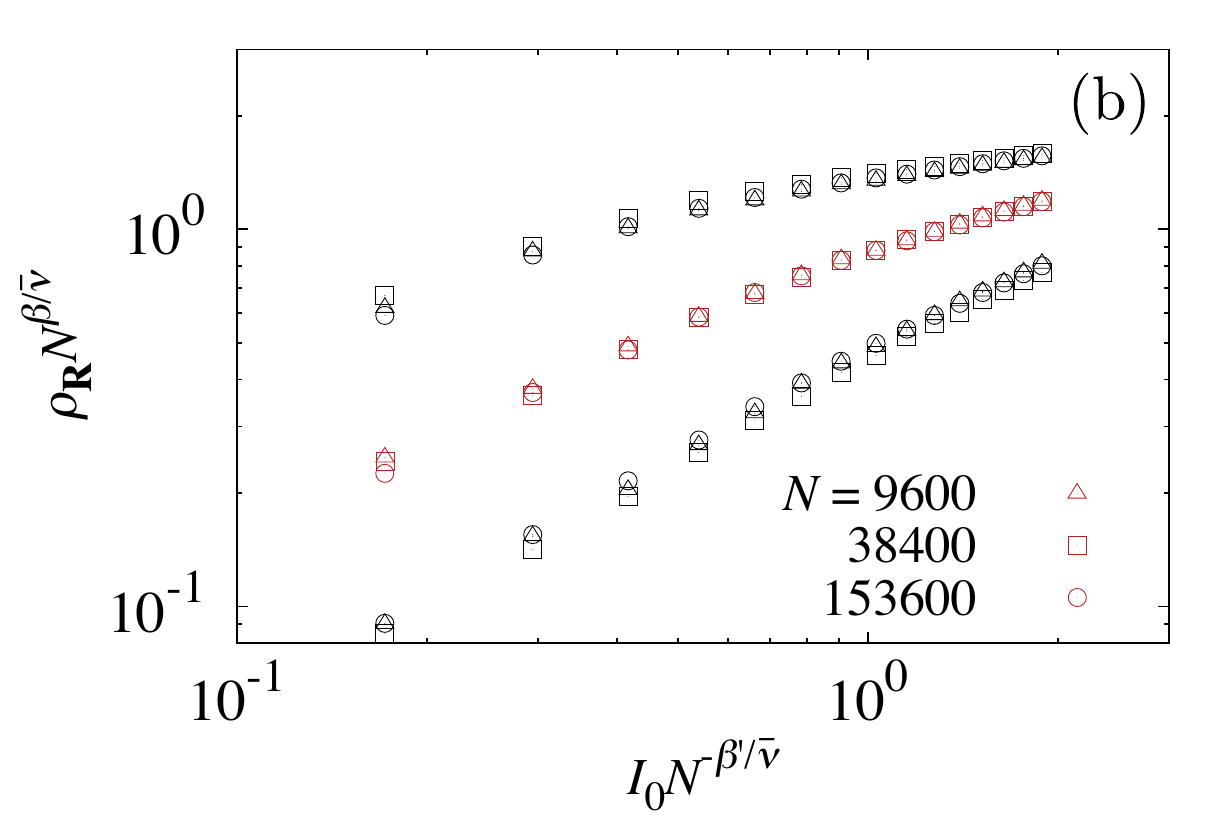}\\
\includegraphics[width=0.85\columnwidth]{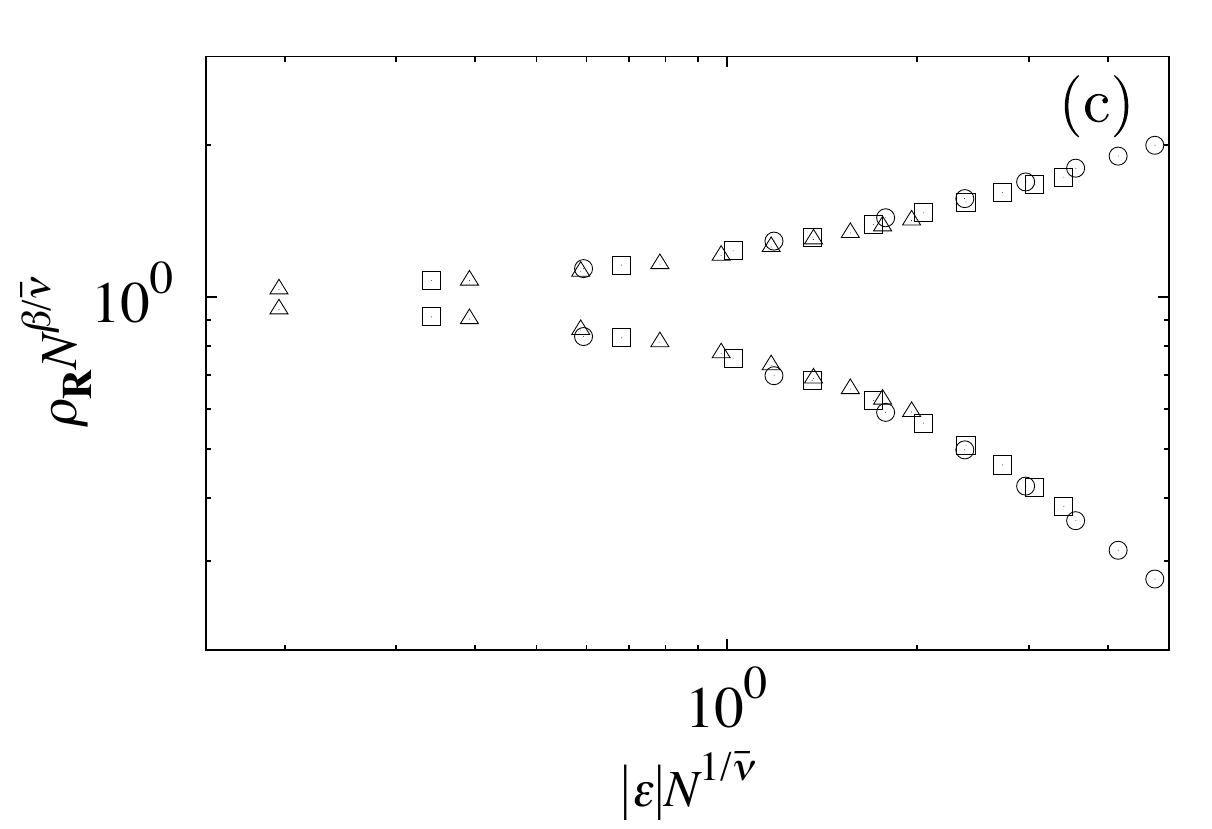}
\caption{\label{fig:fig5} (Color online) Numerical verifications of the static FSS form $\Phi_\rho$ for $\rho_{_\mathbf{R}}$ on Poisson random networks with $\langle k\rangle = 5$ at tricriticality $\mu = \mu_t$. (a) A three-dimensional (3-d) plot of data collapse. (b) Cross sections (marked by thick solid lines) of the 3-d plot for $\varepsilon N^{1/\bar\nu} \approx 2$, $0$, $-2$ (from top to bottom). (c) The cross section of the 3-d plot (marked by a dashed line) at $I_0 N^{-\beta'/\bar\nu} \approx 1.27$. The values of tricritical exponents in Table~\ref{tab} are used.}
\end{figure} 

\subsection{Static FSS at tricriticality}

Since $\beta \neq \beta'$ is expected at the tricritical point, there arises the problem of how to verify each of these two exponents numerically. The main difficulty in addressing this problem lies in the lack of a clear distinction between outbreaks and non-outbreaks at finite $N$, which makes $P_\infty$ and $r$ hard to measure. One possible solution is to plant multiple initial seeds, $I_0$, randomly in the network, so that the initial number of $\mathbf{I}$ nodes is no longer limited to $1$. For $N \gg 1$ and $\varepsilon \ll 1$, all seeds are typically very far apart and grow independent of each other, so the outbreak probability is approximately $I_0 P_\infty$, where $P_\infty$ is the outbreak probability for a single seed. A proper renormalization of the system should keep $I_0 P_\infty$ invariant. From Eq.~\eqref{eq:fss_surv}, we observe that $I_0$ must be rescaled as $b^{y_\mathrm{s}} I_0$. Thus, a static FSS form for the mean density $\rho_{_\mathbf{R}}$ can be written as
\begin{equation}\label{eq:fss_rho}
\rho_{_\mathbf{R}}(N,\varepsilon,I_0) = b^{y'_\rho} \rho_{_\mathbf{R}}(b N, b^{-1/\bar\nu} \varepsilon, b^{\beta'/\bar\nu}I_0).
\end{equation}
In the limit $N \to \infty$, we expect $\rho_{_\mathbf{R}} \simeq r I_0 P_\infty \sim I_0 \varepsilon^{\beta+ \beta'}$ for $\varepsilon \ll 1$, which implies $y'_\rho = \beta/\bar\nu$. Choosing $b = 1/N$, we can reduce Eq.~\eqref{eq:fss_rho} to
\begin{equation} \label{eq:reduced_fss_rho}
\rho_{_\mathbf{R}}(N,\varepsilon,I_0) = N^{-\beta/\bar{\nu}}\Phi_\rho (\varepsilon N^{1/\bar{\nu}}, I_0 N^{-\beta'/\bar{\nu}}).
\end{equation}

The FSS form of Eq.~\eqref{eq:reduced_fss_rho} can be checked by a three-dimensional data collapse, as shown in the left subplot of Fig.~\ref{fig:fig5}. Also see the other two subplots of Fig.~\ref{fig:fig5} for a closer view along two-dimensional cross sections marked by solid (red or gray) and dashed (black) lines. The results confirm the tricritical exponents $\beta$, $\beta'$, and $\bar\nu$ of Table~\ref{tab}.

\section{Summary and outlook} \label{sec:summary}

In summary, we explored the universality classes of the GEP initiated by a single infected seed (or a finite number of them) on Poisson random networks. Based on some simple theoretical arguments, we obtained a complete view of the phase diagram, derived (tri)critical behaviors in the vicinity of continuous transitions, and postulated the multi-variable FSS forms for various quantities, all of which have been numerically verified.

Our results raise a few notable issues. First, we showed that any phase transition of the GEP on Poisson random networks coincides with an ordinary bond percolation transition. This indicates that even discontinuous transitions are accompanied by a divergent correlation volume, which is a property also observed in some equilibrium systems with long-range interactions~\cite{Bar2014a,Bar2014b}. Whether there is a unified description of these equilibrium models and the nonequilibrium GEP is an interesting question.

Second, the critical exponents of the GEP listed in Table~\ref{tab} are quite different from those associated with doubly infected clusters of cooperative co-epidemics~\cite{WCai2015,Grassberger2015}, in spite of many similarities between the models (e.g., both models behave like the ordinary SIR model until the system percolates). The origins of such differences are yet to be identified.

Finally, we can readily apply the methods discussed here to more general kinds of random networks with locally tree-like structures, such as scale-free networks whose structural heterogeneity would certainly affect the phase diagram and critical behaviors in nontrivial ways~\cite{KChungPrep}. Addressing these issues would broaden our knowledge about universal features of contagion under the influence of cooperative effects.

\begin{acknowledgments}
This work was supported by the National Research Foundation of Korea funded by the Korean government
[Grants No. 2014R1A1A4A01003864 (K.C., M.H.) and No. 2011-0028908 (K.C., H.J.)].
Y.B. is supported in part at the Technion by a fellowship from the Lady Davis Foundation.
We also thank Byungnam Kahng for helpful discussions and Daniel Kim for assistance.
\end{acknowledgments}

\appendix

\section{Derivation of exponent $\beta'$} \label{app:beta}

As noted in the main text, the outbreak probability can be calculated from the relation
\begin{equation} \label{eq:p_infty_supp}
P_\infty = 1 - \sum_{R = 0}^\infty P_R.
\end{equation}
The right-hand side can be obtained by a standard generating function technique~\cite{Callaway2000,Newman2001,Newman2002}, which we summarize in the following.

Provided that $R$ is finite, we consider a generating function for the number of $\mathbf{R}$ neighbors of a randomly chosen node, which is given as~\footnote{Generally speaking, one should distinguish between the case when the node is randomly chosen from the case when the node is one end of a randomly chosen link. Poisson random networks are special in that these two cases are equivalent.}
\begin{equation} \label{eq:g}
G_\lambda(x) = \sum_{k = 0}^\infty p_k (1 - \lambda + \lambda x)^k = e^{-\lambda c(1-x)}.
\end{equation}
The absence of $\mu$ here is a consequence of the locally tree-like structure mentioned in the main text. Now we consider a generating function for the distribution $P_R$, which is defined as
\begin{equation}
H_\lambda(x) \equiv \sum_{R = 1}^\infty P_R x^R.
\end{equation}
The function satisfies a self-consistency relation
\begin{align} \label{eq:h_self_consistency}
H_\lambda(x) &= x G_\lambda \left(H_\lambda(x)\right).
\end{align}
From Eqs.~\eqref{eq:p_infty_supp}, \eqref{eq:g}, and \eqref{eq:h_self_consistency}, we obtain
\begin{align}
1 - P_\infty &= G_\lambda(1-P_\infty) = e^{-\lambda cP_\infty}.
\end{align}
An expansion of this relation about $P_\infty = 0$ gives
\begin{align}
(\lambda c - 1) P_\infty - \frac{(\lambda cP_\infty)^2}{2} + O(P_\infty^3) = 0,
\end{align}
which has a nonzero solution for $P_\infty$ if and only if $\lambda \ge \lambda_c \equiv 1/c$. For $\varepsilon \equiv (\lambda - \lambda_c)/\lambda_c \ll 1$, the nonzero solution satisfies
\begin{equation} \label{eq:betap}
P_\infty \sim \varepsilon^{\beta'}, \quad\beta' = 1.
\end{equation}

\begin{figure}[b]
\includegraphics[width=0.925\columnwidth]{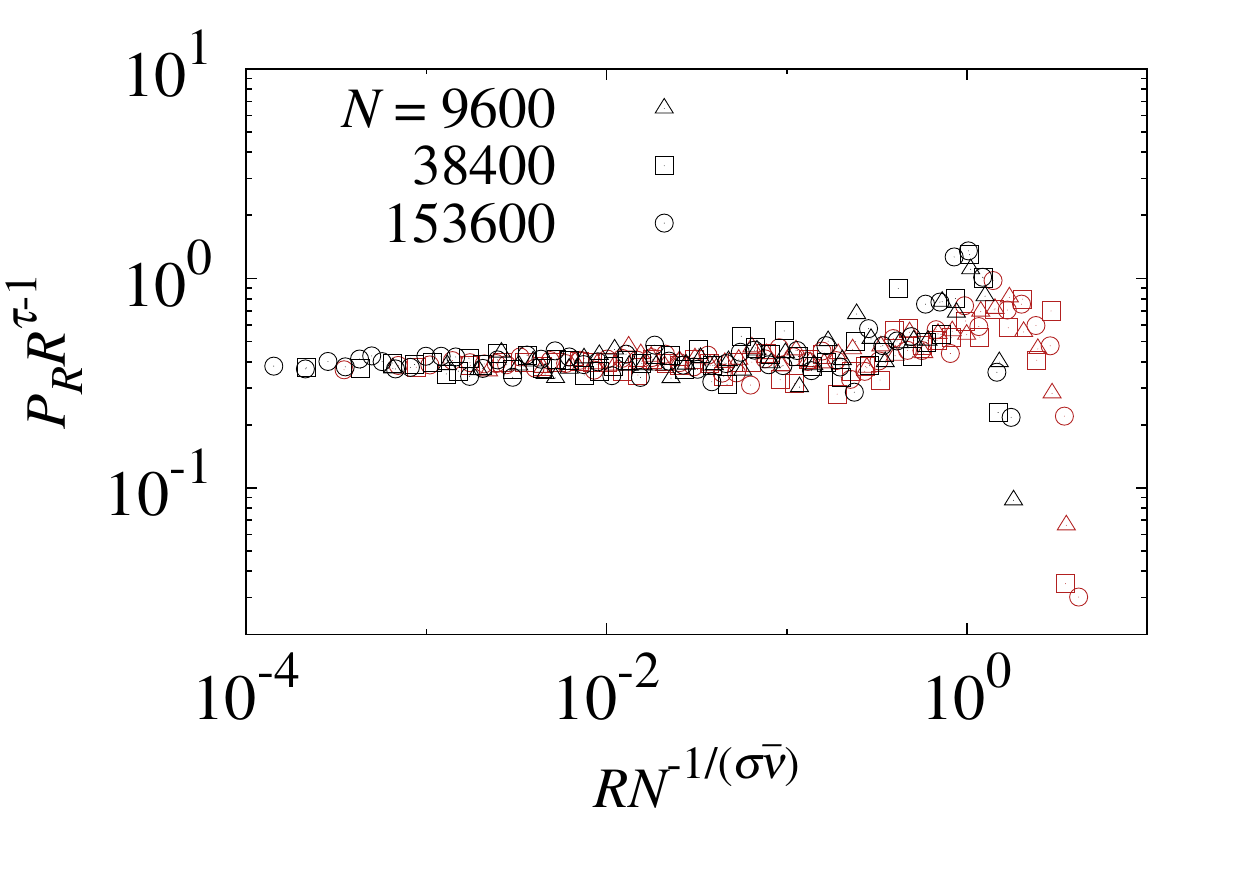}
\caption{\label{fig:figPR} (Color online) The FSS form $\Phi_R$ of the cluster size distribution $P_R$ is numerically verified at the epidemic threshold $\lambda = \lambda_c$ at $\mu = 0.21$ (red or gray) and  $0.25$ (black).}
\end{figure}

\section{Derivation of exponent $\tau$} \label{app:tau}

Using the ansatz $P_R \sim R^{1-\tau}$ at $\lambda = \lambda_c$, the exponent $\tau$ can also be obtained by a generating function technique~\cite{Cohen2002}.
Since $\langle R \rangle$ diverges at criticality, $\tau$ must satisfy $2 < \tau \le 3$. As a result, for $\Delta \ll 1$,the generating function of $P_R$ satisfies 
\begin{equation}\label{eq:h0_tau}
1 - H_{\lambda_c}(1-\Delta) \simeq a \Delta^{\tau - 2}
\end{equation}
where $a$ is a constant. From Eq.~\eqref{eq:h_self_consistency}, we obtain
\begin{equation}
H_{\lambda_c}(1-\Delta) \simeq (1-\Delta) G_{\lambda_c}(1-a\Delta^{\tau-2}),
\end{equation}
whose asymptotic expansion gives
\begin{equation}\label{eq:h0_expansion} 
1 - a\Delta^{\tau-2} \simeq 1 - a\Delta^{\tau-2} - \Delta + \frac{a^2 \Delta^{2(\tau-2)}}{2} + \cdots.
\end{equation}
Since Eq.~\eqref{eq:h_self_consistency} is exact for any $x$, the $\Delta$ term and the $\Delta^{2(\tau-2)}$ term in Eq.~\eqref{eq:h0_expansion} should be able to cancel each other. This gives $\tau=5/2$, which agrees with the MF bond percolation result.

Just like $\beta'$, the value of $\tau$ is unaffected by $\mu$. This is clearly shown in Fig.~\ref{fig:figPR}, which shows data collapses according to a finite-size scaling form
\begin{equation}
P_R \sim R^{1-\tau} e^{-R/R_0(N)} \sim R^{1-\tau}\Phi_R (RN^{-1/\sigma\bar{\nu}}).
\end{equation}

\bibliography{ref-LM15010E-CBHJ}

\end{document}